\documentclass[preprint,preprintnumbers,amsmath,a4paper,aps,12pt,endfloats*]{revtex4}

\usepackage{bm}
\usepackage{graphicx}
 
\newcommand{\BEA}{\begin{eqnarray}}
\newcommand{\BEAN}{\begin{eqnarray*}} 
\newcommand{\EEA}{\end{eqnarray}}
\newcommand{\EEAN}{\end{eqnarray*}}

\begin{document}

\baselineskip 0.85cm


\title{Examination of the Feynman-Hibbs Approach in the Study of
  Ne$_N$-Coronene Clusters at Low Temperatures}

\author{Roc\'{\i}o Rodr{\'\i}guez-Cantano}
\affiliation{Instituto de F\'{\i}sica Fundamental (IFF-CSIC), Serrano 123,
28006 Madrid, Spain}

\author{Ricardo P{\'e}rez de Tudela}
\affiliation{Lehrstuhl f{\"u}r Theoretische Chemie, Ruhr-Universit{\"a}t
  Bochum, 44780 Bochum, Germany} 

\author{Massimiliano Bartolomei}
\affiliation{Instituto de F\'{\i}sica Fundamental (IFF-CSIC), Serrano 123,
28006 Madrid, Spain}

\author{Marta I. Hern\'{a}ndez}
\email{marta@iff.csic.es}
\affiliation{Instituto de F\'{\i}sica Fundamental (IFF-CSIC), Serrano 123,
28006 Madrid, Spain}

\author{Jos\'e Campos-Mart\'{\i}nez}
\affiliation{Instituto de F\'{\i}sica Fundamental (IFF-CSIC), Serrano 123,
28006 Madrid, Spain}

\author{Tom{\'a}s Gonz{\'a}lez-Lezana}
\affiliation{Instituto de F\'{\i}sica Fundamental (IFF-CSIC), Serrano 123,
28006 Madrid, Spain}

\author{Pablo Villarreal}
\affiliation{Instituto de F\'{\i}sica Fundamental (IFF-CSIC), Serrano 123,
28006 Madrid, Spain}

\author{Javier Hern{\'a}ndez-Rojas}
\affiliation{Departamento de F\'{\i}sica and IUdEA,
Universidad de La Laguna, 38203 Tenerife, Spain}

\author{Jos{\'e} Bret{\'o}n}
\affiliation{Departamento de F\'{\i}sica and IUdEA,
Universidad de La Laguna, 38203 Tenerife, Spain}

\date{\today}

\begin{abstract}
Feynman-Hibbs (FH) effective potentials constitute an appealing approach for
investigations of many-body systems at thermal equilibrium since they allow us to
easily include quantum corrections within standard classical simulations. In this
work we apply the FH formulation to the study of Ne$_N$-coronene clusters
($N=$ 1-4, 14) in the 2-14 K temperature range. Quadratic (FH2) and quartic
(FH4) contributions to the effective potentials are built upon Ne-Ne and
Ne-coronene analytical potentials. In particular, a new corrected expression
for the FH4 effective potential is reported. FH2 and FH4 cluster energies and
structures -obtained from energy optimization through a basin-hoping
algorithm as well as classical Monte Carlo simulations- are reported and
compared with reference path integral Monte Carlo calculations. For temperatures $T
> 4$ K, both FH2 and FH4 potentials are able to correct the purely
classical calculations in a consistent way. However, the FH approach fails at
lower temperatures, especially the quartic correction. It is thus crucial to
assess the range of applicability of this formulation and, in particular, to
apply the FH4 potentials with great caution. A simple model of $N$
isotropic harmonic oscillators allows us to propose a means of estimating the
cut-off temperature for the validity of the method, which is found to increase
with the number of atoms adsorbed on the coronene molecule.

\vspace{1cm}

\noindent


\end{abstract}

\maketitle

\newpage

\section{Introduction}

Quantum effects are very important in a variety of many-body systems at
thermal equilibrium, especially for light molecules and/or low
temperatures. Among the problems of interest, we mention the
behavior of water\cite{Paesani:09}, the  
 storage of small molecules in nanoporous
materials\cite{Broom:13,RSCAdv:2012,MnezMesa:12,Nguyen2010} and the study of
molecules trapped inside low temperature matrices, either
solid\cite{Bahou:14,Garkusha:13} or superfluid as in the case of He
nanodroplets\cite{Yang:13,Szalewicz:08}. 
The path-integral formulation of statistical mechanics\cite{Feynman:72} has provided the
framework for the development of accurate path-integral Monte Carlo (PIMC) 
methods\cite{Ceperley:95,Charu:97} to study these systems. However, these methods are
computationally very demanding so various approximate approaches have been
developed over the years to overcome this drawback.

One of the simplest approximations is the use of Feynman-Hibbs (FH) effective
potentials\cite{FHcite} in classical Monte Carlo (CMC) or molecular dynamics
simulations. These potentials are given as a temperature- and mass-dependent
expansion of the intermolecular potentials in powers of $\hbar$. In this way,
the approach provides an easy and appealing means to include quantum
corrections in a purely classical simulation. This formulation has been
applied to the study of both
homogeneous\cite{Guillot:98,Wales:01,Tchouar:04,Abbaspour:14} and heterogeneous
systems as, for instance, the sieving of H$_2$ and D$_2$ in microporous
materials\cite{Nguyen2010,Bhatia:2005,Bhatia:2006,Noguchi:08,Liu:12,Contescu:13}. 
In the case of homogeneous media, the validity of this method has been
investigated in detail by Ses{\'e}\cite{Sese94,Sese95} from comparisons with
``exact'' PIMC calculations of Lennard-Jones systems ($⁴$He, Ne, Ar, D$_2$,
CH$_4$). Similarly, Calvo {\em et   al}\cite{Wales:01} found that the quadratic FH
effective potential reproduces quite well the thermodynamic properties
(melting temperatures, heat capacities, caloric curves, etc.) of Ne, Ar, and
Xe rare gas clusters. More recently, Kowalczyk {\em et al}\cite{Kowalczyk:09}
assessed the FH approach for supercritical $^4$He at 10 K and concluded that
the FH potentials are only suitable at low fluid densities, suggesting that
previous applications to dense para-H$_2$ in nanoporous materials at low
temperatures should be revised. It is thus interesting to investigate the
validity of the FH approach for heterogeneous systems such as fluid-solid
mixtures, for which specific studies are scarcer.

In this work we study the performance of the FH approach for
Ne$_N$-coronene ($N=1-4, 14$) clusters at temperatures ranging from 2 to 14 K, by
comparing CMC calculations using FH potentials with the accurate PIMC
method. This system can be considered as a prototype for studies of van der
Waals interactions between small molecules and carbonaceous
substrates\cite{Marlies:2013} and, in this way, its 
findings may serve as a guide for future simulations of the storage of gases
by new porous carbon materials\cite{max:Carbon2015}. In addition, the
interaction between coronene and other polycyclic aromatic hydrocarbons (PAHs)
with Ne atoms is interesting in connection with the spectroscopy of these
molecules in Neon matrices at low temperatures ($\approx$ 6 K), aimed to
assign some bands observed from various astrophysical
environments\cite{Garkusha:13,Joblin:94,Steglich:10,Garkusha:11}.

The FH effective potentials used in the present work are built upon pairwise
analytical Ne-Ne and Ne-coronene potentials and are obtained both at quadratic
($\hbar^2$) and quartic ($\hbar^4$) order, hereafter referred as FH2 and FH4
potentials, respectively. As in our recent study of He$_N$-coronene
clusters\cite{RguezCantano:15}, minimum energies of the bare interaction
potential are obtained by means of a basin-hopping (BH)
approach\cite{walesd97a}, and the optimized structures are 
then used as seeds for the CMC and PIMC calculations. In this work, we
additionally run BH and CMC calculations with the FH2 and FH4 potentials in
order to assess the extent of improvement with respect to the use of the bare
potentials.

The paper is organized as follows.  Section 2 presents the FH effective
potentials applied to the Ne$_N$-coronene interaction. BH, CMC and PIMC
computational methods are briefly reviewed in Section 3. Results (energies and
structures at different temperatures) are reported and discussed in Section 4.
Finally, concluding remarks are given in Section 5.

\section{Ne$_N$-coronene Interaction Potentials}

\subsection{Bare potentials}

The coronene molecule is assumed to be rigid and fixed to the reference frame.
The origin of the coordinate system is placed at the coronene center of mass,
with the $z$ axis being perpendicular to the molecular plane and the $x$ axis
being overimposed to two of the C-C bonds of this molecule. The position of 
$i-$th Ne atom is given by the Cartesian vector ${\bf r}_i$. 

The total potential of the Ne$_N$-coronene system is given as a sum of
pairwise interactions, 

\vspace{-0.25cm} 

\begin{equation}
V({\bf r}_1,...,{\bf r}_N) = \sum_{i=1}^{N} V_{\rm Ne-Cor}({\bf r}_i) +
\sum_{i<j}^{N} V_{\rm Ne-Ne}(\rho_{ij}), 
\label{NeNCorPot}
\end{equation}

\noindent
where $V_{\rm Ne-Cor}$ and $V_{\rm Ne-Ne}$ are the Ne-coronene and Ne-Ne 
interaction potentials, respectively, and $\rho_{ij} = |{\bf r}_i-{\bf r}_j|$
is the distance between $i-$th and $j-$th Ne atoms.

The Ne-Ne potential is represented by the Improved Lennard-Jones (ILJ)
formula\cite{ILJ}:  

\vspace{-0.65cm} 

\begin{eqnarray}
V_{\rm Ne-Ne}(\rho)  & = & 
\frac{\varepsilon}{m(\rho)-6}  \left[ 6 \left( \frac{\rho}{\rho_e} \right) ^{-m(\rho)}  -
m(\rho)  \left( \frac{\rho}{\rho_e} \right)^{-6} \right],
\label{ILJ1}
\end{eqnarray}

\noindent
where $\varepsilon$ is the well depth, $\rho_e$ is the equilibrium distance, and 

\vspace{-0.45cm} 

\begin{eqnarray}
m(\rho)  & = & \gamma + 4  \left( \frac{\rho}{\rho_e} \right)^2.
\label{ILJ2}
\end{eqnarray}

\noindent
Although Eq. \ref{ILJ1} is more involved than the standard Lennard-Jones
expression, the ILJ potential gives a more realistic representation of both
the size repulsion and the long-range dispersion attraction, as discussed
elsewhere\cite{ILJ}. Values of the parameters $\varepsilon$, $\rho_e$ and
$\gamma$ have been taken from Ref.{\cite{ILJ} and are given in Table
  \ref{TableI}.

The Ne-coronene potential is given as a sum of atom-bond pairwise
contributions\cite{Pirani:04,grapheneours:2013}, 

\vspace{-0.45cm} 

\begin{eqnarray}
V_{\rm Ne-Cor}({\bf r}) & = &  \sum_k U_k(\rho_k,c_k),
\label{abILJ1}
\end{eqnarray}

\noindent
where $k$ runs for the number of bonds (C-C and C-H) in coronene, $\rho_k$ is
the distance between the atom and the bond center and $c_k= \cos(\theta_k)$,
$\theta_k$ being the Jacobian angle describing the orientation of Ne relative
to the bond axis. Both $\rho_k$ and $c_k$ are functions of the Ne position
(${\bf r}$) as well as of the location and orientation of the bond
$k$. Geometrical parameters of coronene are as in
Ref. \cite{grapheneours:2013}. The atom-bond pair potential $U_k$ is
represented, as in Ne-Ne, by an ILJ formula (Eq.\ref{ILJ1}), with the
special feature that the well depth and the equilibrium distance vary
with $c_k$ as\cite{Pirani:04}

\vspace{-0.45cm} 

\begin{eqnarray}
\varepsilon (c_k) & = & \varepsilon^{\perp} + 
                      \left( \varepsilon^{\parallel}-\varepsilon^{\perp}
                      \right) \, c_k^2 \nonumber\\
\rho_e (c_k) & = & \rho_e^{\perp} + \left( \rho_e^{\parallel}-\rho_e^{\perp} \right) \, c_k^2,
\label{abILJ2}
\end{eqnarray}

\noindent
where  $\varepsilon^{\perp}$ ($\varepsilon^{\parallel}$) and $\rho_e^{\perp}$
($\rho_e^{\parallel}$) are the well depth and equilibrium distance for the
perpendicular (parallel) orientation of the atom with  
respect to the bond. These parameters were initially estimated from
 the polarizability of the interacting partners, and subsequently fine-tuned
 from the comparison with benchmark high level {\em ab initio} calculations
 involving large basis sets\cite{grapheneours:2013}. Their values are
 listed in Table \ref{TableI} for the two types of the involved bonds, C-C and C-H. 

It is worth noting that the atom-bond additive representation takes into
consideration three-body effects since it explicitly makes use of the
components of the bond polarizability tensor and in the end it provides a good
estimation of 
the total (coronene) molecular polarizability. On the other hand, the present
model neglects three-body effects arising between two rare gas atoms and a
third body (either Ne or coronene), which however
are expected to be negligible for the investigated equilibrium geometries.

\subsection{Feynman-Hibbs effective potentials}

The FH effective potentials for an A-B interacting system are obtained
from the following gaussian average of the $V_{\rm AB}$ potential
\cite{FHcite,Sese94}  

\vspace{-0.35cm} 

\begin{eqnarray}
V_{\rm AB}^{\rm FH}(\bf r) & = & \left( \frac{6 \mu}{\pi \hbar^2 \beta} \right)^{3/2} 
                  \int d{\bf u} \; \; V_{\rm AB}({\bf r} + {\bf u}) \; 
                  \exp\left(-\frac{6 \mu}{\beta \hbar^2} u^2 \right),
\label{FH1}
\end{eqnarray}

\noindent
where ${\bf r}$ is the vector joining A and B, $\mu$ is the A-B reduced mass,
and $\beta= 1/k_BT$, $k_B$ being the Boltzmann constant.  Eq.\ref{FH1} is
based on a variational treatment of the path-integral and involves the neglect
of exchange effects\cite{Sese94}. 
Closed expressions of the FH potential are obtained upon a Taylor expansion of 
$V_{\rm AB}({\bf r} + {\bf u})$ 
around the vector ${\bf r}$ 
in powers of the Cartesian components of ${\bf u}$  
and solving the corresponding integrals of Eq.\ref{FH1}. We have\cite{Sese95}

\begin{eqnarray}
V_{\rm AB}^{{\rm FH}(2p)}(\bf r) & = & \sum_{n=0}^p \frac{1}{n!} \left(
\frac{\beta \hbar^2}{24 \mu} \right)^{n}  
                 \nabla^{2n} \left[ V_{\rm AB}({\bf r}) \right], 
\label{FH1b}
\end{eqnarray}

\noindent
where $\nabla^{0}=1$ and $p=0, 1, 2$ correspond to the bare, quadratic (FH2)
and quartic (FH4) potentials, respectively. Note that 
the integral of Eq. \ref{FH1} vanishes for the contributions involving odd
powers of the components of ${\bf u}$. It can be seen that at high
temperatures the effective potentials tend to the bare potential and the
classical regime is correctly recovered. However, as temperature tends to zero
these potentials become unphysically high, which will inevitably impose limits to
the model.

In the case of the Ne-Ne interaction, $V_{\rm AB}({\bf r})$ depends on $|{\bf
  r}|\equiv\rho$ and the FH2 potential becomes 

\begin{eqnarray}
V_{\rm Ne-Ne}^{\rm FH2}(\rho) & = & V_{\rm Ne-Ne}(\rho) \,
 + \, \frac{\hbar^2 \beta}{24 \mu} 
\left( \, V_{\rm Ne-Ne}''(\rho) + \frac{2}{\rho}V_{\rm Ne-Ne}'(\rho) \, \right)  , 
\label{FH2}
\end{eqnarray}

\noindent
where $V_{\rm Ne-Ne}$ is the potential of Eq.\ref{ILJ1} and $V_{\rm Ne-Ne}'$
and $V_{\rm Ne-Ne}''$ are their first and second derivatives, respectively,
with respect to $\rho$.

The quartic Ne-Ne effective potential (FH4) writes 

\vspace{-0.5cm} 

\begin{eqnarray}
V_{\rm Ne-Ne}^{\rm FH4}(\rho) & = & V_{\rm Ne-Ne}^{\rm FH2} + \frac{1}{2} \left(
\frac{\hbar^2 \beta}{24 \mu} \right)^2 \left( \, V_{\rm Ne-Ne}''''(\rho) +
\frac{4}{\rho}V_{\rm Ne-Ne}'''(\rho) \, \right),  
\label{FH3}
\end{eqnarray}

\noindent
where, analogously, $V_{\rm Ne-Ne}'''$ and $V_{\rm Ne-Ne}''''$ are the third
and fourth derivatives of $V_{\rm Ne-Ne}$ with respect to $\rho$. 

It is worth pointing our that Eq. \ref{FH3} differs from previously published
expressions\cite{Bhatia:2005,Bhatia:2006,Kowalczyk:09,Liu:12}, which we believe
are incorrect. More details are provided in the Appendix.

The form of the Ne-coronene FH2 potential is more complicated because 
 the atom-bond potentials $U_k$  of Eq.\ref{abILJ1} depend both on the
 distance to the bond center $\rho_k$ and the cosine of the Jacobian angle,
 $c_k$. The result is 

\vspace{-0.35cm}

\begin{eqnarray}
V_{\rm Ne-Cor}^{\rm FH2}({\bf r}) & = & V_{\rm Ne-Cor}({\bf r}) \nonumber \\
 & &  + 
\frac{\hbar^2 \beta}{24 m_{\rm Ne}} {\Huge \sum_k} \left\{ \, \frac{\partial^2
  U_{k}}{\partial \rho_k^2}  
+ \frac{2}{\rho_k}  \frac{\partial U_{k}}  {\partial \rho_k} 
+ \frac{(1-c_k^2)}{\rho_k^2} \frac{\partial^2 U_{k}}{\partial c_k^2}
- \frac{2 c_k}{\rho_k^2} \frac{\partial U_{k}}  {\partial c_k }
\, \right\}.  
\label{FH4}
\end{eqnarray}

\noindent
Note that in Eq.\ref{FH4} 
the Ne mass is written instead of the reduced mass of Ne-coronene. We have
made this choice because the coronene molecule is fixed to the reference frame
and it is thus assumed that it has an infinite mass. 

We have found that the contribution to $V_{\rm Ne-Cor}^{\rm FH2}$ due 
to the derivatives of $U_k$ with respect to $c_k$ are very small. To simplify
the calculations, we have assumed that the contributions from these 
derivatives to the terms of $\hbar^4$ order are negligible. In this way, the
FH4 potential is written as   

\vspace{-0.35cm}

\begin{eqnarray}
V_{\rm Ne-Cor}^{\rm FH4}({\bf r}) & = & V_{\rm Ne-Cor}^{\rm FH2}({\bf r}) 
+ \frac{1}{2} \left( \frac{\hbar^2 \beta}{24 m_{\rm Ne}} \right)^2 
{\Huge \sum_k} \left\{ \, \frac{\partial^4 U_{k}}{\partial \rho_k^4} 
+       \frac{4}{\rho_k}  \frac{\partial^3 U_{k}}  {\partial \rho_k^3} \, \right\}. 
\label{FH5}
\end{eqnarray}

\noindent 
Nevertheless, the exact expression for the correction of $\hbar^4$ order is
given in Eq. \ref{nabla4uk}, and it has been additionally checked that the
contribution to that equation of the derivatives of the atom-bond potential
with respect to $c_k$  is negligible. 

More details about the derivation of Eqs. \ref{FH2}-\ref{FH5} are provided in
the Appendix.  All the required derivatives of the ILJ functions were obtained
analytically as well as the gradients of the potentials of
Eqs.\ref{FH2}-\ref{FH5} (needed for the application of the BH algorithm
referred below). In any case, recent advances in techniques of automatic
differentiation\cite{Griewank:2008} might prove advantageous for this kind of
simulations.  

The Ne-Ne bare potential as well as the FH2 and FH4 ones at 6 K are presented 
in Fig. \ref{fig1}. The ILJ form of the bare potential is very realistic, as
we have compared it with the Tang-Toennies potential\cite{TangToennies:03} and
found that both potentials  
would appear as indistinguishable in Fig. \ref{fig1}. Indeed, the well depth and
equilibrium distance of the Tang-Toennies potential are 3.646 meV and 3.090
\AA, very close to the values of Table \ref{TableI}. On the other hand, it can be
seen in Fig. \ref{fig1} that the FH corrections significantly modify the bare
potential, the effective potentials being more ``repulsive'': the equilibrium
distance changes from 3.09 to 3.24 (3.28) \AA $\;$ and the well depth, from 3.66
to 3.03 (2.84) meV, as one goes from the bare to the FH2 (FH4) potentials,
respectively.    

The corresponding potentials for Ne-coronene are shown in
Fig.\ref{fig2} as functions of the $y$ coordinate, while $z$ and $x$ are fixed
at the absolute minimum of the classical potential (at 3.21 and 0 \AA,
respectively). Again, it can be seen that the effect of the quadratic
correction is non negligible, for instance, the minimum energy moves from
-27.83 (bare potential) to -25.68 meV (FH2 potential) but, on the other hand, the
FH4 potential is very close to the FH2 one.

\section{Classical and Quantum-Mechanical Calculations}

\subsection{General procedure and notation}

In this work we compare the performance of the Ne$_N$-coronene FH2 and FH4
effective potentials with that of the bare interaction potentials. 
First, we will study the equilibrium geometries of these clusters by means of the
BH approach. The corresponding calculations are denoted by BH,
BH-FH2 and BH-FH4 for the bare, FH2 and FH4 potentials, respectively. Note
that, while the BH calculations are independent of the temperature, BH-FH2 and
BH-FH4 must be repeated for the different temperatures of the study. The BH 
 equilibrium geometries are used as initial configurations of the
CMC calculations at each temperature, which will be denoted as CMC, CMC-FH2
and CMC-FH4, for the bare, FH2 and FH4 potentials, respectively. The resulting
energies and configurations are compared with the PIMC calculations,
where just the bare interaction potential is employed. 

For the BH calculations using the bare interaction potential, the zero point
energy (ZPE) was also computed within the harmonic approximation and added to
the BH equilibrium energies (details can be found in
Ref.\cite{RguezCantano:15}). This is a convenient estimation of the quantum
effects of the system when the thermal effects become small, as will be
discussed in Section IV. These calculations are denoted by BH+ZPE.

\subsection{BH minimization}

Likely candidates for the global potential energy minima of Ne$_N$-coronene
clusters were located using the BH scheme \cite{walesd97a}, which is also
known as the ``Monte Carlo plus energy minimization'' approach of Li and
Scheraga \cite{lis87}. This method transforms the potential energy surface
into a collection of basins and explore them by hopping between 
local minima. This technique has been used successfully for both neutral 
\cite{walesd97a,rojas2012,silvia2012} and charged atomic and molecular
clusters \cite{rojas,wales2,wales3,rojas1,rojas2,rojas2010}, along with many
other applications.\cite{wales4} In the size range considered here the global
optimization problem is feasible at a reasonable computational cost. 
A total of 5 runs of $5\times 10^4$ BH steps each were performed for all
clusters sizes. The global minimum was generally found in fewer than
10$\text{\textsuperscript{4}}$ BH steps. The optimization temperature was
chosen between 8 and 10 K. 

\subsection{PIMC and CMC calculations}

Details of the PIMC method employed here can be found in our study on
He$_N$-coronene clusters \cite{RguezCantano:15} and in previous 
literature\cite{Ceperley:95,KCW:JCP96,RLPGDVG:CTC12,RPLGGDV:EPJD13,RGVG:JCP15}. 
The basic assumption involves expressing the density matrix at a
temperature $T$ as a product of $M$ density matrices at higher temperatures $MT$:

\begin{equation}
\label{ec_rho}
\rho({\cal R}_0,{\cal R}_{M}; \beta)= \int d{\cal R}_1 \textellipsis
d{\cal R}_{M-1} \prod_{\alpha=0}^{M-1} 
\rho({\cal R}_{\alpha},{\cal R}_{\alpha+1};\eta) ,
\end{equation}
\noindent
where $\eta = \beta/M$. ${\cal R}_{\alpha}$ is the vector which collects the
$3N$ positions of the $N$ Ne atoms: ${\cal R}_{\alpha} \equiv \{ {\bf
  r}_1^{\alpha}, \textellipsis, {\bf r}_N^{\alpha} \}$, being {\bf
  r}$_i^{\alpha}$ the position vector of the $i-$th Ne atom at the time {\it
  slice} or {\it imaginary} time $\alpha$. The total Hamiltonian $\hat{H}$ of
the system with the coronene molecule fixed to the origin of coordinates can
be written as: 

\begin{equation}
\label{hamiltonian}
 \hat{H} = 
-\frac{\hbar^{2}}{2m_{\rm Ne}}\sum_{i=1}^N\nabla_{i}^{2} + V({\cal R}).
\end{equation}
\noindent 
where $V$ is the interaction potential of Eq.\ref{NeNCorPot}.

The internal energy is obtained by means of the virial estimator
\cite{HBB:JCP82,GF:JCP02} as:

\begin{eqnarray}
\label{Et}
\langle E(T) \rangle\! & = &  
{3 N \over 2 \beta} - \left \langle \! 
{1 \over 2M} \sum_{\alpha=0}^{M-1} \sum_{i=1}^N 
({\bf r}_i^{\alpha}\!-\!{\bf r}_i^{\rm C}) 
\cdot {\bf F}_i^{\alpha} 
 - \!{1 \over M} \sum_{\alpha=0}^{M-1}  
V({\cal R}_{\alpha}) 
\! \right \rangle.
\end{eqnarray}

\noindent
where ${\bf r}_i^{\rm C}= M^{-1} \sum_{\alpha=0}^{M-1} {\bf r}_i^{\alpha}$ is
the centroid of the $i$th particle and ${\bf F}_i^{\alpha}$ is the force
experienced by the $i$ particle on the $\alpha$ slice. The integration is
carried out via a Metropolis Monte Carlo algorithm, as an average over a
number of paths $\{{\cal R}_1,{\cal R}_2,\textellipsis, {\cal R}_{M},{\cal
  R}_{M+1}\}$ sampled according to a probability density proportional 
to the factorized product of $M$ density matrices of Eq.\ref{ec_rho}. Exchange
effects are neglected. The number of beads vary between $M$= 1 (for the CMC,
CMC-FH2 and CMC-FH4 calculations) to a maximum of 150 for the lowest
temperature PIMC simulations. Depending on $M$, the number
of steps varies between 10$^5$ to 10$^7$. The staging sampling method has been
employed\cite{chandler:85} involving a number of eight beads in each movement
for the PIMC simulations. The final average energy is obtained by
extrapolation to the $M \rightarrow \infty$ following a parabolic
law\cite{Roy:06,PerezTudela:11}.

\section{Results and Discussion}

\subsection{Cluster energies and structures at 6 K}

Before tackling the study of the FH approximation as a function of the
temperature, we start presenting results at 6 K, an intermediate value in the
range addressed here and in coincidence with the temperature of various
experiments on PAHs isolated in Ne matrices\cite{Garkusha:13,Garkusha:11,Steglich:10}.   

In Table \ref{TableII}, energies of various Ne$_N$-coronene clusters, $N$=1-4
and 14, at 6 K as obtained from the BH and CMC approaches and using the bare,
FH2 and FH4 effective potentials are reported and compared with the PIMC
results.  Various arrangements (or ``isomers'') $(n_a,n_b)$ are examined for
each number of Ne atoms, where $n_a$ and $n_b$ refer to the number of atoms
placed above and below the coronene plane, respectively. It can be seen that
both the BH and the CMC energies tend to the PIMC energies as one goes from
using the bare to the FH2 and FH4 potentials. However, given the difference
between the BH and the CMC energies, it is clear that thermal effects are non
negligible at this temperature so, among all the calculations, the CMC-FH4
energies are the ones giving the closest agreement with the reference PIMC
results. In addition, notice that CMC-FH4 agrees with PIMC as to which is the
most stable isomer $(n_a,n_b)$ for a given cluster size $N$. For example, for
$N=4$, the most stable arrangement within the FH4 potential is the (3,1) one,
in agreement with the PIMC calculation, whereas (4,0) gives the absolute
minimum when using the bare and  FH2 potentials. This result can be explained
by a more repulsive character of Ne-Ne interaction when it is considered at the
FH4 level (Fig.\ref{fig1}), thus making an arrangement with a larger density
of Ne atoms relatively less stable (as the (4,0) one).  

Also from Table \ref{TableII}, it is worth mentioning that the addition of the
ZPE to the BH energies gives a fair agreement with the PIMC energies, and that
the BH+ZPE calculations correctly predict the relative stability of the
different isomers for a given $N$.

Computation of the energies per atom helps us to quantify the performance of
the FH approach. Results are depicted in Fig.\ref{fig3}, where it is
clear the improvement of the CMC-FH energies with respect to the CMC
ones. 
In more detail, note that, for a given $N$, the differences between the CMC
and PIMC energies are larger for those clusters having a larger number of
atoms on a given side of the molecule. In this case, quantum effects increase
because the number of effective Ne-Ne interactions increases as well 
(interactions between atoms sitting on different sides of the molecule are
negligible). For example, for (3,0), with three effective Ne-Ne interactions,
the relative error of the CMC calculation is 14 \%, whereas for the (2,1)
cluster, with just one Ne-Ne interaction, the error reduces to 11 \%. The FH2
and FH4 potentials certainly amend the classical result, although the errors
are also larger when the number of interacting atoms increase. Indeed, the
errors of the CMC-FH2 calculation are 5 \% and 2 \% for the (3,0) and (2,1)
clusters, respectively. It is important to mention that both Ne-Ne and
Ne-coronene interactions contribute significantly to the quantum corrections:  
for the (3,0) cluster, 24 \% and 76 \% of the FH2 correction (computed at the
corresponding optimal geometry) are due to Ne-Ne and
Ne-coronene, respectively, whereas these numbers become 12 \% and 88\% in the
case of the (2,1) arrangement.

Besides increasing the overall energies, there are some changes in the 
geometries of the Ne$_N$-coronene clusters when the interaction potentials are
modified. Probability densities as functions of the coordinates
parallel to the coronene plane, ${\cal D}(x,y)$, as obtained from PIMC,
CMC-FH4 and CMC, are shown in Fig.\ref{fig4} for (7,7) Ne$_{14}$-coronene at 6 
K (results for the FH2 potential are not shown as they are quite close to the
FH4 ones). These plots have been obtained by means of a histogramming
procedure on the $x$ and $y$ coordinates, accumulating the probability density
along the $z$ coordinate. For each side of the molecule, one
Ne atom is placed above the central hollow, while the other six atoms are located
near the borders of the outer hexagons. The distributions from the different
calculations are quite similar. However, it can be noticed that the peaks of
the PIMC and CMC-FH4  probability densities are somewhat wider and that the
average distance between those peaks is slightly larger than in the case of
the CMC calculation. These features can be further examined by means of the
one-dimensional distribution ${\cal D} \left ([x^2+y^2]^{1/2} \right)$ as
shown in Fig.\ref{fig5} (it is computed by accumulating the probability density
${\cal D}(x,y)$ along the angle $\varphi$ given by $\tan{\varphi} = y/x$). The
CMC-FH4 distribution is in better agreement with the PIMC calculations than
the purely classical one. Fig.\ref{fig5} also depicts the distance of the
outer Ne atoms with respect to the coronene symmetry axis ($z$) as obtained
from the BH and BH-FH4 optimized geometries, and it 
can be seen that these positions coincide with the maxima of the corresponding
CMC distributions. The FH4 potential involves larger equilibrium distances than
the bare potential (for instance, see Fig. \ref{fig1}), a feature that can
explain the shift in the peak position of the CMC-FH4 with respect to the CMC
one. The peak of the PIMC distribution is in the middle, suggesting that 
the FH4 potential is overestimating this effect.

\subsection{Temperature dependence of the cluster energies}

Further insight is gained into the FH approach by studying the cluster 
 energies as functions of the temperature for the different cluster sizes. In
 Fig.\ref{fig6}, the behavior of the (1-4, 0) and (7,7) cluster energies as
 obtained from the different potentials and methods is tested against PIMC in
 the temperature range 2-14 K. To make a more coherent  comparison, the
 energies have been shifted by the minimum energy (BH) and 
 divided by the number of atoms, 
   
\begin{equation}
{\widetilde E}(T) = \frac{E(T)- E^{\rm BH}}{N}.
\label{enerfig6}
\end{equation}

\noindent
Although not all of the ($n_a$=1-4,0) arrangements correspond to the
absolute minimum energy (see Table \ref{TableII}), we have chosen this
sequence in order to study the FH approach as a function of the number of
Ne atoms over a given face of the molecule. 

In the higher temperature range of Fig. \ref{fig6} ($T >$ 6 K) the three CMC
calculations give a monotonous increase of the
cluster energies with the temperature, in agreement with the behavior of the
PIMC energies. In this region, the CMC-FH2 method considerably improves the
cluster energies with respect to the CMC calculations, while the CMC-FH4
calculation just adds a small correction to the CMC-FH2 results. This
is illustrated in Table \ref{TableIII}, where the relative errors
of the three CMC calculations are listed at $T= 10$ K. Note that ``un-shifted''
energies were taken for the calculation of the relative errors (i.e., $E$
instead of ${\widetilde E}$). Note that the CMC errors rise 
with the number of Ne atoms, as already discussed above. It can also be seen
from the Table that the CMC-FH2 and
CMC-FH4 calculations roughly halve the CMC errors independently of the number
of Ne atoms. Therefore, although the FH2 and FH4 potentials do
not provide a perfect agreement with the PIMC energies, they do 
introduce quantum corrections in a steady way. 

As temperature decreases ($T \lesssim $ 6 K), quantum effects become
dominant over the thermal ones and a different behavior of the methods is
evident. On the one hand, the slope of the PIMC energies is modified to reach
a horizontal asymptote given by the ZPE of the system. 
The CMC energies, on the other hand, tend (linearly) to zero, deviating
considerably from the quantum calculation. Finally, the CMC-FH2 and CMC-FH4
energies reach a minimum at about 6 K and rapidly increase as temperature
decreases, also in contrast with the correct behavior given by PIMC. In this
way, the effective potentials are not adequate for temperatures below
$\lesssim$ 4 K.

The functional dependence of the CMC-FH2 (CMC-FH4) energies with the
temperature can be easily understood as soon as it is realized that they are almost
equal to the sum of the BH-FH2 (BH-FH4) and the CMC energies, which account
for quantum and thermal effects, respectively. In other words, in the present
system, the behavior of the CMC-FH2 (CMC-FH4) energies is identical
to that of CMC except for the (temperature-dependent) modification of the local minimum
energy due to the effective FH2 (FH4) potentials. 

It is also worthwhile to note that the slope of all the CMC curves of
Fig. \ref{fig6} is roughly 3
$k_B$, in agreement with a model of $N$ classical harmonic oscillators in a
three-dimensional (3D) space. In addition, the PIMC energies at  low
temperatures are in a fairly good agreement with the ZPE computed here within
the harmonic approximation (dotted lines in Fig. \ref{fig6}). In this way, we have tested
a model of $N$ 3D isotropic harmonic oscillators against present
calculations. It is assumed that each of the $N$ Ne atoms moves under an
effective potential $V_{\rm har}(r) = m_{\rm    Ne} \omega_N r^2 /2$, where
$\omega_N$ is a characteristic frequency which varies with $N$. Its value is
obtained from equating the computed ZPE per atom to $3 \hbar \omega_N /2$. In
this way, the ``average interaction'' undergone by each atom -which depends
on the interaction with both the molecular substrate and the remaining atoms
in the cluster- is approximated by an isotropic harmonic potential. Within
this model, the quantum mechanical average energy per atom is
\cite{Tuckerman:10}   

\vspace*{-0.6cm}

\begin{eqnarray}  
{\widetilde E}^{\rm q}_{\rm har} (T) = \frac{3}{2} \, \hbar \omega_N \, \coth{\frac{\hbar
    \omega_N}{ 2 k_B T}},
\end{eqnarray}
 
\noindent
to be compared with the PIMC energy. Applying Eq. \ref{FH1b} to the harmonic
potential $V_{\rm har}$, the FH2 potential is 

\vspace*{-0.6cm}

\begin{eqnarray}  
{\widetilde E}^{\rm BH-FH2}_{\rm har} (T) = \frac{1}{8} \frac{(\hbar \omega_N)^2}{k_B T}, 
\end{eqnarray}

\noindent
which will be associated with the BH-FH2 energy. Note that the FH4 potential is
identical to the FH2 one within the present harmonic approximation. The FH2
thermal energy per atom is obtained by adding $3 k_B T$,

\vspace*{-0.6cm}

\begin{eqnarray}  
{\widetilde E}^{\rm CMC-FH2}_{\rm har} (T) = {\widetilde E}^{\rm BH-FH2}_{\rm
  har} (T) + 3 k_B T, 
\end{eqnarray}

\noindent
which will be related to the CMC-FH2 energies. Results are shown in
Fig. \ref{fig7} for the (2,0) and (4,0) clusters. Although anisotropy and
anharmonicity effects should probably be added to attain a more quantitative
agreement, the model energies compare fairly well with the computed ones and,
in this way, this simple model does provide an adequate zero-order
description of the behavior of these clusters.

\subsection{Applicability of the FH2 and FH4 effective potentials}

The results of Fig. \ref{fig6} indicate that both FH2 and FH4 potentials
are useful for $T >$ 4 K. It is evident, however, that below this
temperature the CMC energies computed with these potentials deviate
considerably from the reference PIMC behavior, and that this effect is more
pronounced for the quartic (FH4) potential. This is in accord with the
indication of Ref. \cite{Tchouar:04} that the FH2 approximation is valid whenever the
next term in the expansion (the quartic one) remains much smaller than the
quadratic term. In a more detailed study of Ne
and $^4$He Lennard-Jones systems, Ses{\'e}\cite{Sese95} found that the
FH2 potential generally performs better than the FH4 one. In this work, we have
found that the FH4 potential is useful just in a narrow range
of temperatures (around 6 K), since at larger temperatures its
corrections become negligible whereas at lower temperatures it worsens the FH2
estimations.  

It is thus important to determine the temperature range of validity of these
effective potentials as they can lead to erroneous results when applied below
a critical temperature $T^{\star}$. Ses{\'e} has found that, for relatively
low densities, this temperature can be deduced from the condition
$ \lambda_B /\sigma < 0.5$, where $\lambda_B=  \hbar \sqrt{2 \pi / m
  k_B T^{\star}}$ is the thermal de Broglie wavelength and $\sigma$
is the Lennard-Jones collision diameter. Here, taking $\sigma$= 2.76
\AA \hspace*{0.1cm} (the diameter of the Ne-Ne ILJ potential), we obtain
$T^{\star} \approx 8$ K. This is a somewhat strict condition in view of the
specific results of this work. 

Based in the model of isotropic harmonic oscillators described above, we have
found a simple means of estimating the critical temperature $T^{\star}$ that
might be extrapolated to related systems where PIMC calculations would
become too time-consuming. This temperature is defined as the
crossing between the approximate (${\widetilde E}^{\rm CMC-FH2}_{\rm har}$)
and the quantum (${\widetilde E}^{\rm q}_{\rm har}$) energies. Given that at low 
temperatures ${\widetilde E}^{\rm q}_{\rm har}(T) \approx \frac{3}{2} \hbar \omega_N $,
the ZPE per atom of the system, a simple quadratic equation follows, whose 
lowest root is

\begin{equation}
T^{\star} = \frac{3}{2} \hbar \omega_N \, \left( \frac{ 1 - \sqrt{1/3}}{ 6
 \, k_B} \right).
\label{last}
\end{equation}

\noindent
The resulting temperatures range from $T^{\star}=$ 2.8 to 4.8 K as we go from
the (1,0) to the (7,7) cluster, respectively. It is worth noting that this model correctly 
takes into account the shift of $T^{\star}$ with the number of rare gas
atoms (density), as found for other systems\cite{Sese94,Kowalczyk:09}. Indeed,
$T^{\star}$ is proportional to the ZPE per atom, which increases with
the number of Ne atoms, as can be seen in Fig. \ref{fig6}. This trend is
possibly due to a rise of the frequencies of the average interaction undergone
by each Ne atom as the number of rare gas atoms surrounding it increases.

\section{Concluding remarks}

The Feynman-Hibbs (FH) approach has been applied to the study of
Ne$_N$-coronene clusters ($N=$ 1-4, 14) at low temperatures ($T=$ 2-14 K) and 
using realistic analytical potentials. The suitability of the quadratic
(FH2) and quartic (FH4) effective potentials has been investigated by
comparing basin-hoping (BH) optimizations and classical Monte Carlo (CMC)
calculations of cluster energies and structures with benchmark path-integral
Monte Carlo (PIMC) calculations.  

For $T >$ 4 K it is found that, although there is not a perfect agreement
with the PIMC calculations, the effective potentials significantly improve the
purely classical calculations. Quantum effects -which are significant and
 due to both Ne-Ne and Ne-coronene interactions- are partially corrected by
 the FH potentials in a reliable way. In particular, the FH4
approach at 6 K correctly predicts the most stable structure over a set of
energetically close local minima, and tends to emulate the PIMC probability
distributions (although these distributions do not vary as much as the energies). 

For lower temperatures $T \lesssim$ 4 K, where zero-point energy (ZPE) effects dominate over
thermal ones, the FH formulation fails to reproduce the dependence of the PIMC
energies with the temperature, while the BH+ZPE approximation does
reproduce well the PIMC results. In particular, the FH4 
potential, which generally improves the results of the FH2 potentials at higher
temperatures, deviates more dramatically from the correct results than the FH2
approach. Therefore, the quartic effective potential, which has been recently 
employed in simulations of the diffusion of light molecules in nanoporous
materials\cite{Bhatia:2005,Liu:12,Contescu:13}, should be applied with extreme 
caution. 

We believe that further investigations of the performance of these effective
potentials are worthwhile as they allow us to include quantum effects
(i.e., ZPE effects) into classical simulations in a
straightforward manner. For instance, it would be interesting to explore in
 detail the performance of these potentials for dynamical processes -such as
 the transmission of atoms through nanoporous membranes\cite{jpcaours:2015}-  
using  Path Integral Molecular Dynamics approaches as benchmark quantum simulations.

\section{Appendix: Derivation of the Feynman-Hibbs effective potentials}

In this paragraph we give a more detailed account of the derivation of the
quadratic (FH2) and quartic (FH4) effective potentials for the interaction
potentials of this study (Eqs. \ref{FH2}-\ref{FH5}). Since these potentials
are of the type $V(\rho)$ or $V(\rho,\cos{\theta})$, it is convenient to
consider the operators of Eq.\ref{FH1b} in spherical
coordinates. In particular, the Laplacian is given, 
in terms of ${\bf r} \equiv (r,\cos{\theta},\phi)$, as

\begin{eqnarray}
\nabla^2 & = & \frac{\partial^2 \;}{\partial r^2} + \frac{2}{r}
\frac{\partial \;}{\partial r} + 
\frac{1}{r^2} \left(  (1 - c^2) \frac{\partial^2 \;}{\partial c^2} - 2 c  
\frac{\partial \;}{\partial c} +
\frac{1}{(1 - c^2)} \frac{\partial^2 \;}{\partial \phi^2} \right),
\label{Ap1}
\end{eqnarray} 

\noindent
where $c= \cos{\theta}$.

To apply Eq.\ref{FH1b}, we use the property of invariance of the
Laplacian operator under any translation or rotation in the three-dimensional
space. In the case of the Ne-Ne interaction, the origin of the coordinate
system is displaced to coincide with one of the atoms, and, since $V_{\rm Ne-Ne}$
only depends on the radial distance $\rho$,

\begin{eqnarray}
\nabla^2 V_{\rm Ne-Ne} & = & \frac{d^2 V_{\rm Ne-Ne}}{d \rho^2} + \frac{2}{\rho}
\frac{d V_{\rm Ne-Ne}}{d \rho}, 
\end{eqnarray}

\noindent
leading to the FH2 correction of Eq.\ref{FH2}.

For the Ne-coronene interaction, taking into account Eq. \ref{abILJ1}

\begin{eqnarray}
\nabla^2 V_{\rm Ne-Cor}({\bf r}) & = & \sum_k \nabla^2  U_k(\rho_k,c_k).
\end{eqnarray}

For every bond $k$, $\nabla^2  U_k$ can be easily performed after 
a transformation from the original Cartesian system to a new one where the
origin is at the bond center and the $z$ axis is aligned 
with the bond axis. In this way,  $\rho_k$ and $c_k$ coincide with the radial
distance and the cosine of the polar angle of the new reference system and, 
taking into account Eq.\ref{Ap1} and the independence of the potential with
respect to the azimuthal angle,

\begin{eqnarray}
\nabla^2  U_k & = & \frac{\partial^2
  U_{k}}{\partial \rho_k^2}  
+ \frac{2}{\rho_k}  \frac{\partial U_{k}}  {\partial \rho_k} 
+ \frac{1}{\rho_k^2} \left( (1-c_k^2) \frac{\partial^2 U_{k}}{\partial c_k^2}
- 2 c_k \frac{\partial U_{k}}  {\partial c_k } \right),
\end{eqnarray}
    
\noindent
which can be readily related to Eq. \ref{FH4}.

To obtain the FH4 potential of the Ne-Ne interaction, the $\nabla^4$ operator
from Eq. \ref{FH1b} is applied to $V_{\rm Ne-Ne}$ as

\begin{eqnarray}
\nabla^4 V _{\rm Ne-Ne}& = & \left( \frac{d^2 }{d \rho^2} + \frac{2}{\rho}
\frac{d }{d \rho} \right)  
\left( \frac{d^2 V_{\rm Ne-Ne}} {d \rho^2} + \frac{2}{\rho}
\frac{d V_{\rm Ne-Ne}}{d \rho} \right) \nonumber \\
  & = &  \frac{d^4 V_{\rm Ne-Ne}} {d \rho^4} + \frac{4}{\rho}
\frac{d^3 V_{\rm Ne-Ne}}{d \rho^3}, 
\end{eqnarray}  

\noindent
an expression which can be immediately identified with the FH4 corrections of
Eq.\ref{FH3}.

At this point, it is worthwhile noting that other
authors\cite{Bhatia:2005,Bhatia:2006,Kowalczyk:09,Liu:12} have reported an  extra
term, especifically $\frac{15}{\rho^3}\frac{d V}{d \rho}$, in the formal
expressions of the FH4 potentials (Eq. 2 of
Ref. \cite{Bhatia:2005} for example). We have checked that
the contributions depending on other derivatives of the potential out of the
third and fourth derivatives are canceled out in the
calculation of $\nabla^4 V$, and therefore, that
the correct FH4 potential expression is that given by Eq. \ref{FH3} of
this work.

Finally, the $\nabla^4$ operator applied to the atom-bond potential
$U_k(\rho_k,c_k)$ is

\begin{eqnarray}
\nabla^4 U _{k}& =  &  \frac{\partial ^4 U_{k}} {\partial \rho_k^4} + \frac{4}{\rho_k}
\frac{\partial^3 V_{k}}{\partial \rho_k^3} + \frac{2}{ \rho_k^2} \left[
(1-c_k^2) \frac{\partial^4 U_k}{\partial \rho_k^2 \partial c_k^2} - 2 c_k
\frac{\partial^3 U_k}{\partial \rho_k^2 \partial c_k} \right] \nonumber \\
 & & + \frac{1}{\rho_k^4} \left[ (1-c_k^2)^2 \frac{\partial^4 U_k}{\partial c_k^4}
  - 8 c_k (1-c_k^2) \frac{\partial^3 U_k}{\partial c_k^3} - 4 (1-3c_k^2)
  \frac{\partial^2 U_k}{\partial c_k^2} \right].
\label{nabla4uk}
\end{eqnarray}   

\noindent
Eq. \ref{FH5} is readily obtained once all the derivatives with respect $c_k$
are neglected, as already discussed elsewhere.

\section*{Acknowledgments}

The work has been funded by Spanish MINECO grants FIS2013-48275-C2-1-P,
FIS2014-51993-P, and FIS2013-41532-P. R. R.-C. thanks support from
FIS2013-48275-C2-1-P. Allocation of computing time by CESGA (Spain) and
support by the COST-CMTS Action CM1405 ``Molecules in Motion (MOLIM)'' 
are also acknowledged.


\newpage
\begin{table}
\small
\caption{Parameters of the Ne-Ne, Ne-CC and Ne-CH pair
  potentials\cite{ILJ,grapheneours:2013}.  
Distances are in \AA, energies in meV, and $\gamma$ is dimensionless.}
\label{TableI}
\vspace*{.5cm}
\begin{tabular}{ccccccc}         
\hline
\hline
 & & & &  & & \\
Pair & & $\rho_e$ & & $\varepsilon$ & & $\gamma$ \\
\hline
  & & & & & &   \\
  Ne-Ne & & 3.094  & &   3.660  & & 9.0 \\
  & & & & & &   \\
 
   & & $\rho_e^{\perp}$ & $\rho_e^{\parallel}$   & $\varepsilon^{\perp}$ & $\varepsilon^{\parallel}$ & $\gamma$ \\
 \hline
  & & & & & &   \\
  Ne-CC & & 1.297  & 1.809 &  3.643  & 3.995 & 8.5 \\
  Ne-CH & & 2.544  & 1.782 &  3.316  & 3.655 & 9.0 \\
  & & & & & &   \\
\hline
\hline
\end{tabular}
\end{table}

\vspace{1.5cm}

\begin{table}
\small
\caption{ Ne$_N$-coronene energies (in meV) at 6 K for the different
  calculations of this work (see text for details).  Various isomers $(n_a,n_b)$ are studied, where
  $n_a$ ($n_b$) is the number of Ne atoms above (below) the coronene
  plane. PIMC error bars (in meV), associated to the $M
  \rightarrow \infty$ extrapolation procedure, are given in
  parenthesis. Standard deviation of the CMC energies (not shown) is about
  0.01 meV. For a given $N$, the absolute minimum energy within each method is
  shown in boldface.}
\label{TableII}
\vspace*{.5cm}
\begin{tabular}{cccccccccc}         
\hline
\hline
 & & & &  & & & & & \\
$N$ & ($n_a,n_b$) & BH & BH+ZPE & BH-FH2 & BH-FH4 & CMC & CMC-FH2 & CMC-FH4 & PIMC \\
 & & & &  & & & & & \\
\hline
 & & & &  & & & & & \\
1 & (1,0) & -27.83 & -24.35  & -26.04 & -25.80 & -25.99 & -24.19 & -23.95 &
-23.89 (0.01) \\
 & & & &  & & & &  &\\
2 & (1,1) & {\bf -55.72} & {\bf -48.75} & {\bf -52.12} & {\bf -51.65} & -52.04
& {\bf -48.44} & {\bf -47.96} & {\bf -47.85} (0.10) \\
 & (2,0)  & -55.36       & -47.26 & -51.13       & -50.44       & {\bf -52.20} &	-48.10       & -47.42	& -46.72 (0.03)\\
 & & & &  & & & &  &\\
3  & (3,0) & {\bf -87.42} & {\bf -73.85} & {\bf -80.51} & {\bf -79.23} & {\bf
  -82.44} & {\bf -75.49} & {\bf -74.20}	& {\bf -72.10} (0.01) \\
 & (2,1) & -83.28       & -71.69  & -77.25       & -76.32       & -78.28
&	-72.36       & -71.44	& -70.62 (0.01) \\
 & & & &  & & & &  &\\
4 & (4,0) & {\bf -117.29} & -97.03 & {\bf -106.68} & -104.58	 & {\bf -110.50} & {\bf -99.97} & -97.88 & -94.81 (0.10)\\ 
  & (3,1) & -115.38       & {\bf -98.32}    & -106.66      & {\bf -105.14} &
-108.60       & -99.80       & {\bf -98.28} & {\bf -96.04} (0.20) \\
  & (2,2) & -110.90       & -94.66     & -102.41       & -101.01       &
-104.54       & -96.34       & -94.96 & -93.50 (0.20) \\
& & & &  & & & &  &\\
14  & (7,7)   & -415.15 & -332.53 & -369.99 & -360.34 &  -392.46 & -346.83 & -336.94 & -322.44   (0.02)\\ 

\hline
\hline
\end{tabular}
\end{table}

\begin{table}
\small
\caption{Relative errors (in \%) of the CMC energies with respect to the PIMC
  ones ($\mid E^{\rm CMC}-E^{\rm PIMC} \mid /E^{\rm PIMC}$) as obtained from the different
  potentials and clusters ($n_a,n_b$).  }
\label{TableIII}
\vspace*{.5cm}
\begin{tabular}{ccrrrrr}         
\hline
\hline
 & & & &  & & \\
$T$ (K) & Potential & (1,0) & (2,0) & (3,0) & (4,0) & (7,7) \\
\hline
   & & & & & &   \\
   & Bare & 7 & 10  &  12 &  14  & 20 \\
10 & FH2  & 2 &  4  &   6 &   7  & 10 \\
   & FH4  & 2 &  4  &   5 &   6  &  8 \\
  & & & & & &   \\
\hline
\hline
\end{tabular}
\end{table} 

\newpage

\begin{figure}[h]
\vspace*{-4.4cm}
\hspace*{-.5cm}\includegraphics[width=18cm,angle=0.]{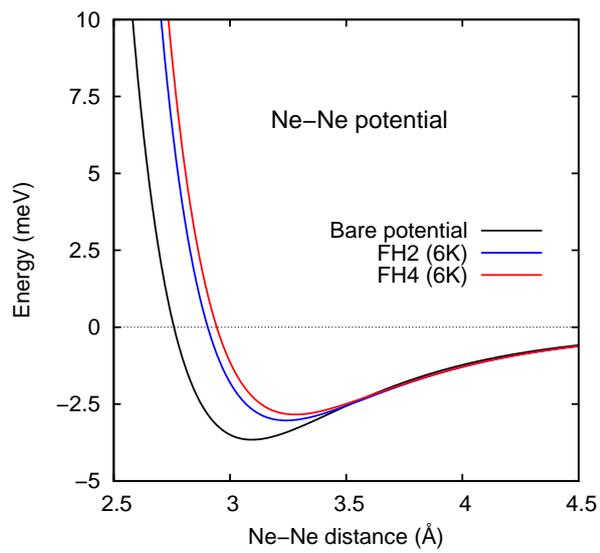}
\caption[] {Bare Ne-Ne interaction potential (meV) as a
  function of the interatomic distance (in \AA), compared with the quadratic
  and quartic Feynman-Hibbs effective potentials at 6 K.}
\label{fig1}
\end{figure}

\begin{figure}[h]
\vspace*{-4.4cm}
\hspace*{-.5cm}\includegraphics[width=18cm,angle=0.]{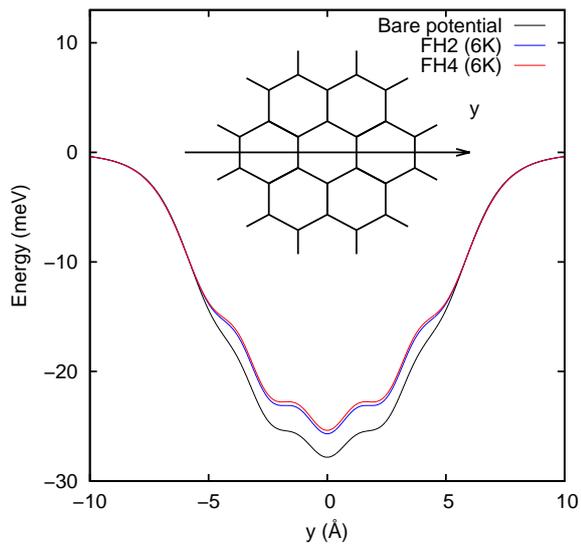}
\caption[] {Ne-coronene (Bare, FH2 and FH4 at 6K) interaction potentials (in meV), 
as functions of the $y$ Cartesian coordinate 
(depicted in the inset), where the $x$ and $z$ coordinates are fixed at the absolute minimum of the 
bare potential (0. and 3.21 \AA, respectively).}
\label{fig2}
\end{figure}

\newpage

\begin{figure}[h]
\vspace*{.4cm}
\hspace*{2.cm}\includegraphics[width=13cm,angle=0.]{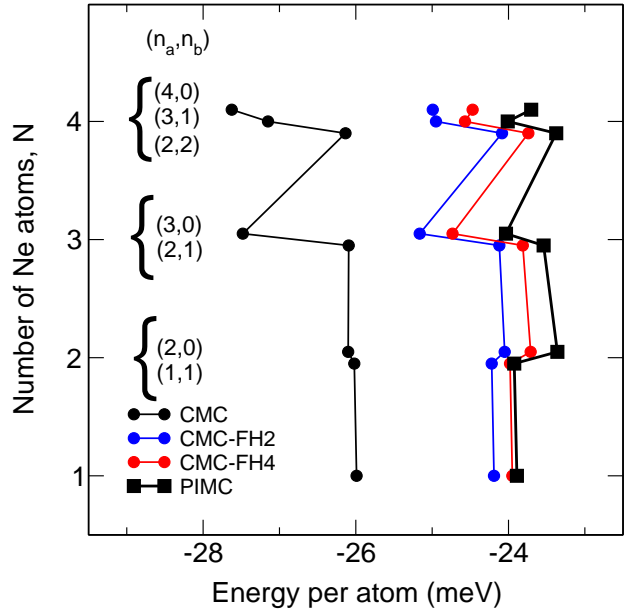}
\caption[] {Energies per atom (in abscissas) of Ne$_N$-coronene at 6 K as
  obtained from CMC, CMC-FH2, CMC-FH4 and PIMC, for the different
  sizes $N$ and isomers $(n_a,n_b)$ (in ordinates).}
\label{fig3}
\end{figure}

\newpage

\begin{figure}[h]
\hspace*{0.2cm}\includegraphics[width=17cm,angle=0.]{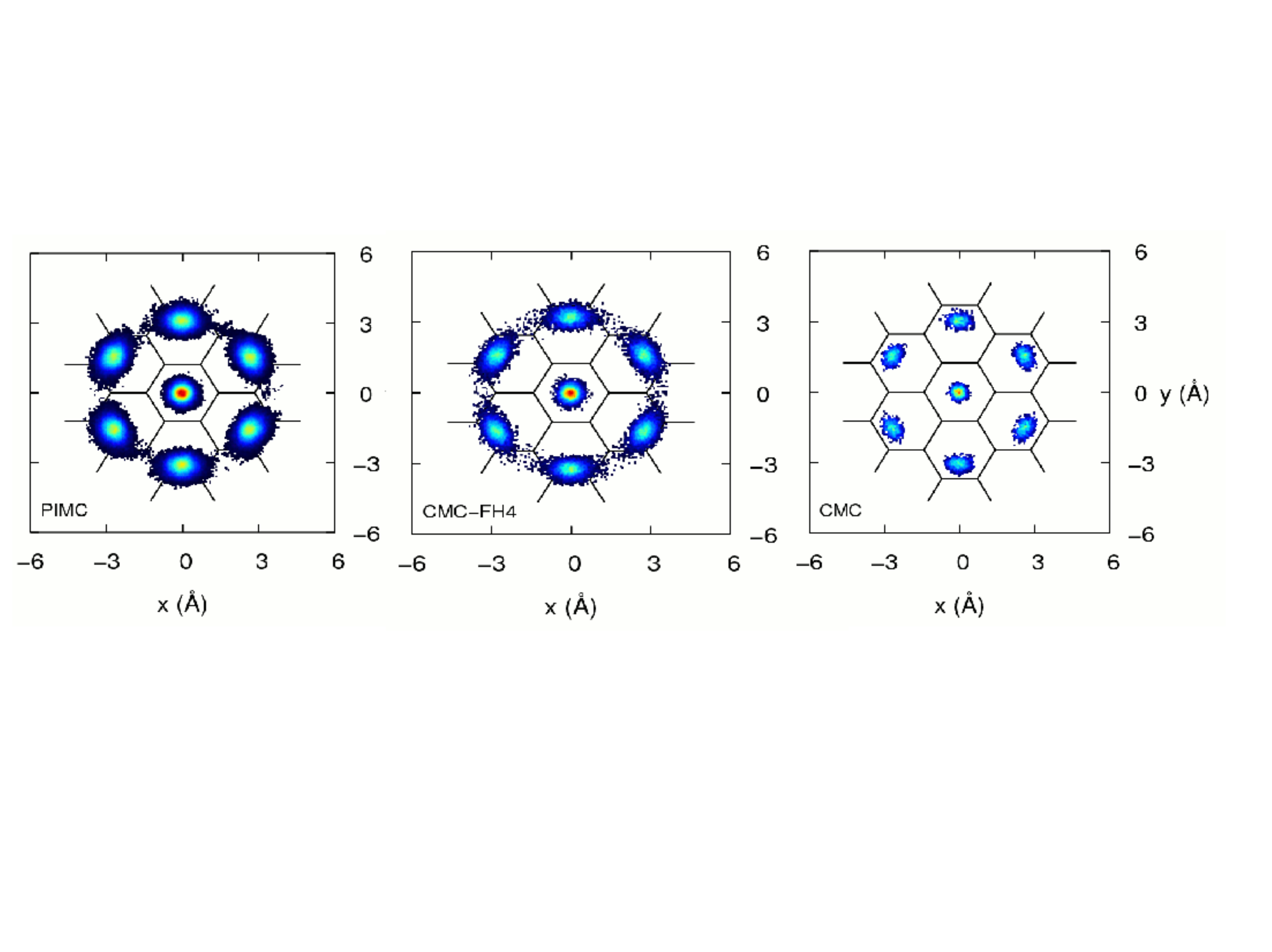}
\caption[] {Probability densities ${\cal D}(x,y)$ of (7,7) Ne$_{14}$-coronene at
  6 K. Left, middle and right panels: PIMC, CMC-FH4 and CMC calculations,
  respectively.}
\label{fig4}
\end{figure}

\newpage 

\begin{figure}[h]
\hspace*{.5cm}\includegraphics[width=15cm,angle=0.]{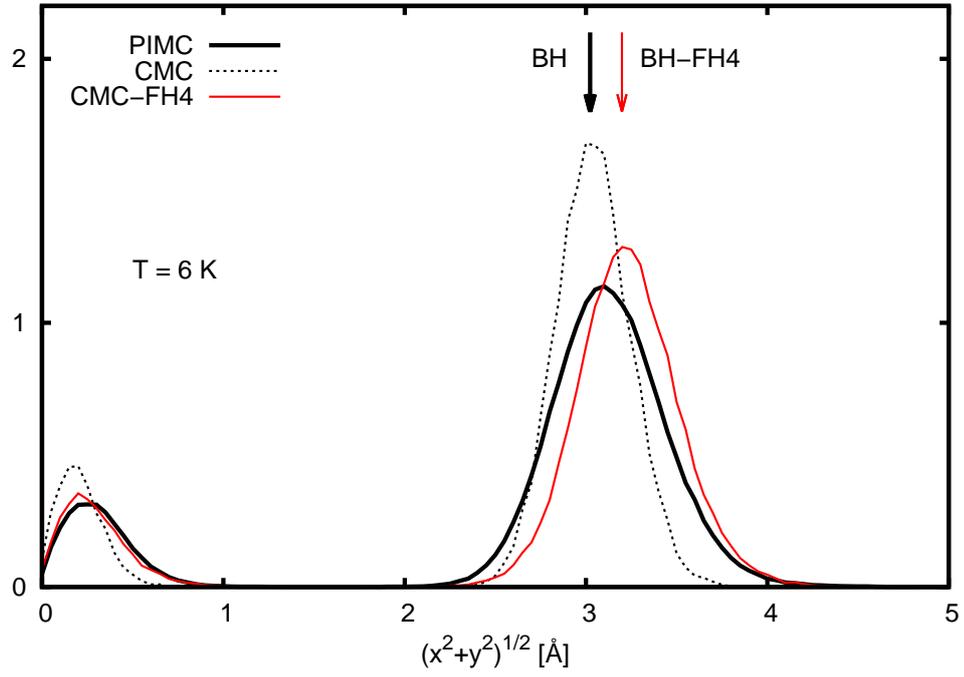}
\caption[] {Normalized probability density ${\cal D} \left( [x^2+y^2]^{1/2} \right)$
   of (7,7) Ne$_{14}$-coronene at 6 K, for CMC, CMC-FH4 and PIMC
  calculations. In addition, the distance of the outer Ne atoms to the
  coronene symmetry axis ($z$), as obtained from the BH and BH-FH4 calculations,
  are depicted by arrows.} 
\label{fig5}
\end{figure}

\newpage 

\begin{figure}[h]
\vspace*{.4cm}
\hspace*{-.5cm}\includegraphics[width=16cm,angle=0.]{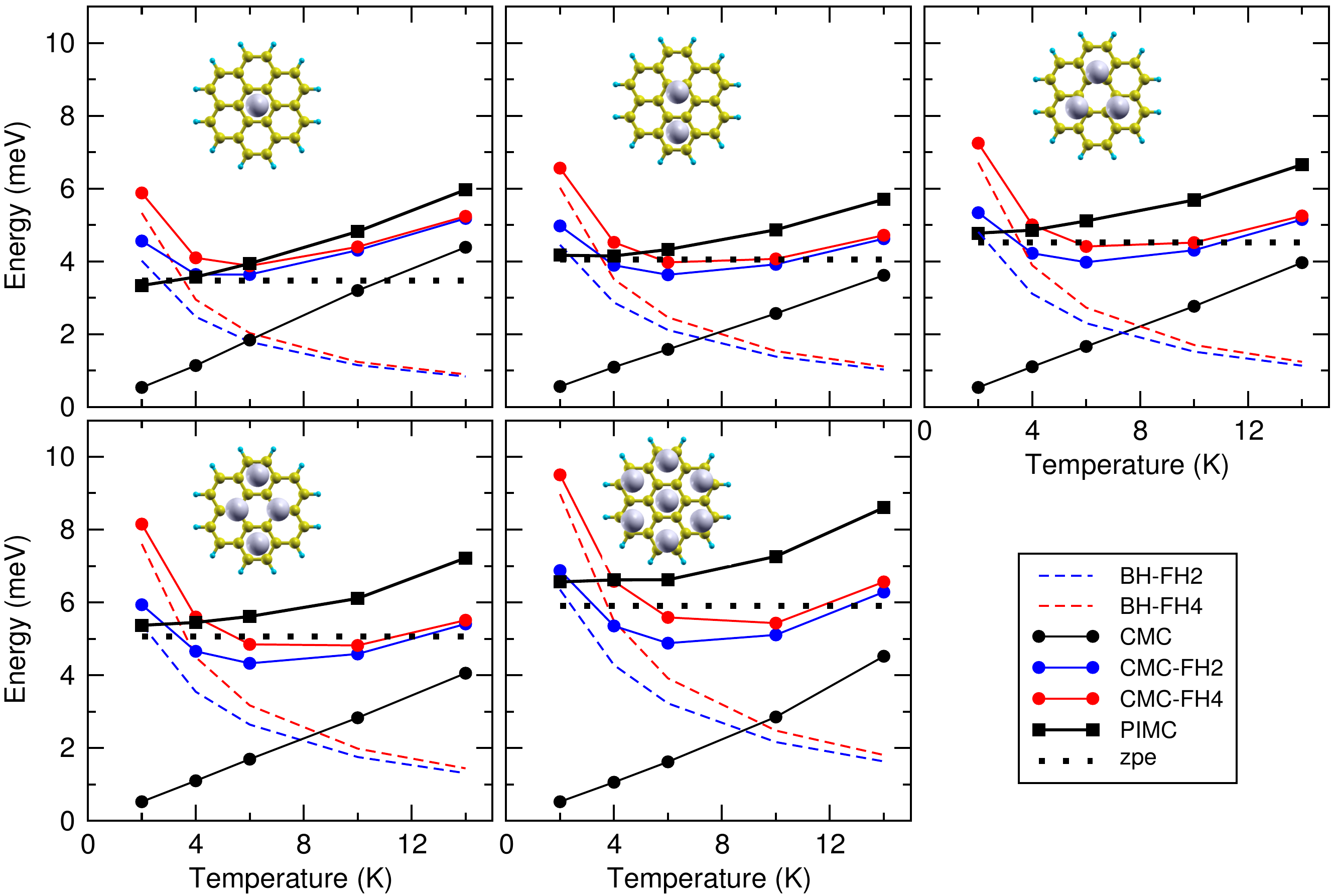}
\caption[] {Energies per atom, shifted as in Eq.\ref{enerfig6}, of
  Ne$_N$-coronene clusters as functions of
  temperature for the different methods used in this work. Insets depict the
  classical optimal geometries of the clusters. (1,0), (2,0), (3,0), (4,0) and
  (7,7) are shown in the left upper, middle upper, right upper, left lower and
  middle lower panels, respectively. Calculations have been performed at 2, 4,
  6, 10, and 14 K (lines are guides to the eye). See text for discussion.} 
\label{fig6}
\end{figure}

\begin{figure}[h]
\vspace*{.4cm}
\hspace*{-.5cm}\includegraphics[width=9cm,angle=0.]{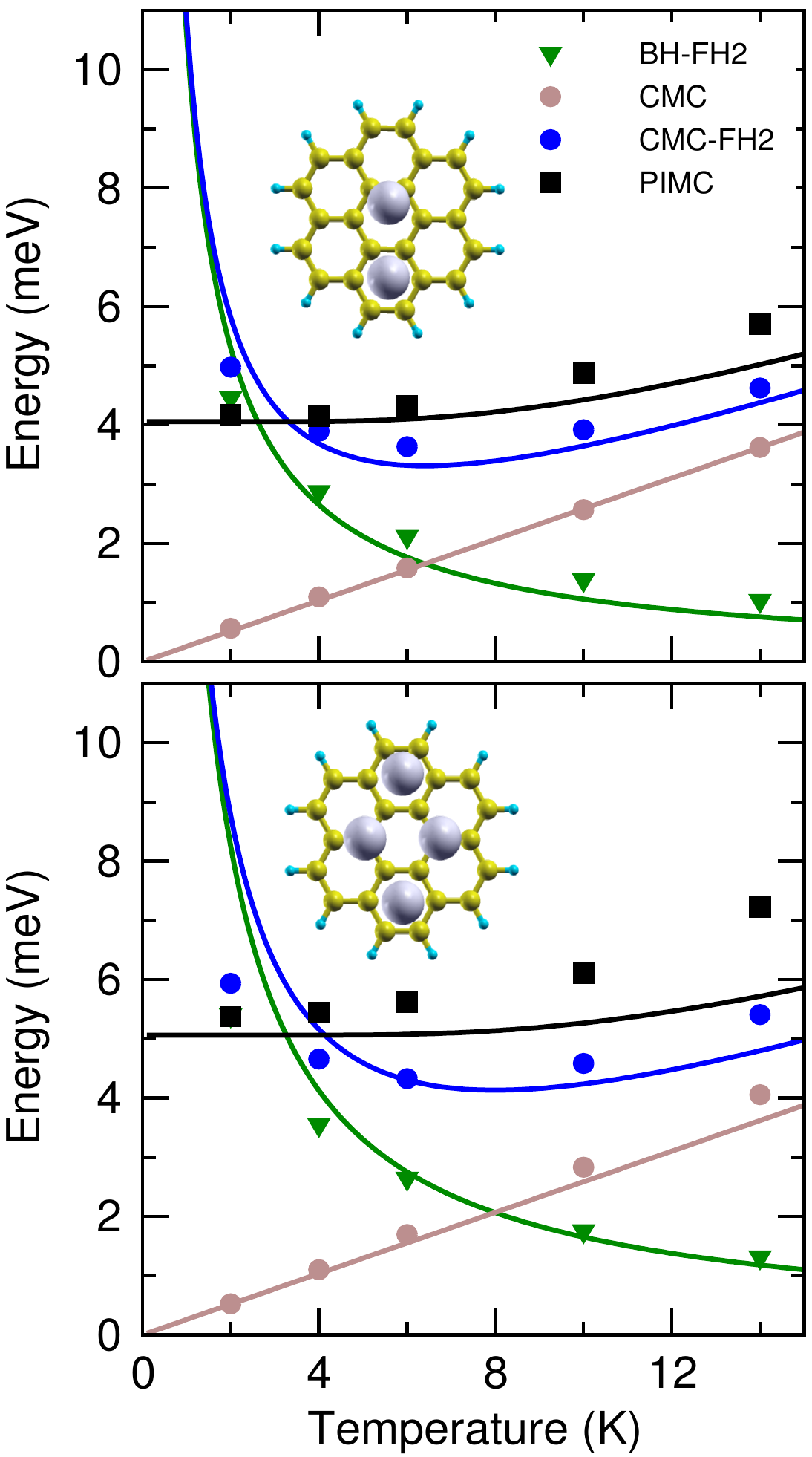}
\caption[] {Comparison between the computed BH-FH2, CMC, CMC-FH2 and PIMC energies
  (symbols) and the corresponding energies obtained with a model of N harmonic oscillators (solid
  lines). Upper and lower panels: (2,0) and (4,0) clusters, respectively. See
  text for details.}  
\label{fig7}
\end{figure}

\end{document}